\documentclass[twocolumn,showpacs,amsfonts,aps,prc,floatfix,superscriptaddress]{revtex4-1}

\usepackage{amsmath}
\usepackage{bm}
\usepackage{graphicx}

\begin{document}

\title{Dissipative Hydrodynamic Effects on Baryon Stopping}

\author{Akihiko Monnai}
\email[]{monnai@nt.phys.s.u-tokyo.ac.jp}
\affiliation{Department of Physics, The University of Tokyo, Tokyo 113-0033, Japan}
\affiliation{Theoretical Research Division, Nishina Center, RIKEN, Wako 351-0198, Japan}

\date{\today}

\begin{abstract}
The quark-gluon plasma is considered to behave as a relativistic viscous fluid in the high-energy heavy ion collisions. In this study, I develop and estimate a second order dissipative hydrodynamic model at finite baryon density with effects of baryon dissipation together with those of shear and bulk viscosities. It is found that the hydrodynamic evolution effectively reduces baryon stopping, suggesting that the collisions are less transparent at the initial stage. Also the net baryon distribution is found sensitive to baryon dissipation as well as to viscosities. The results indicate that the dissipative hydrodynamic modeling would be important for understanding unique properties of the hot medium even in the high-energy collisions.
\end{abstract}


\maketitle

\section{Introduction}
\label{sec1}
\vspace*{-2mm}

The determination of the properties of the quantum chromodynamics (QCD) matter in wider temperature and chemical potential regions has been one of the most important goals in the hadron physics.
The quark matter in high temperature with low baryon chemical potential is experimentally accessible through the high-energy heavy ion collisions, which contributes significantly to the exploration of the QCD phase diagram. 
The early experiments predate to the Alternating Gradient Synchrotron (AGS) in the Brookhaven National Laboratory (BNL) and the Super Proton Synchrotron (SPS) at the European Organization for Nuclear Research (CERN). A sufficiently large amount of energy for the production of the quark-gluon plasma (QGP) \cite{Yagi:2005yb}, a deconfined state of quarks and gluons, is considered to be available in the $\sqrt{s_{NN}} = 200$ GeV Au-Au collisions at the Relativistic Heavy Ion Collider (RHIC) in the BNL \cite{Adare:2008ab}, which is one of the biggest achievements in the hadron physics.
With the beginning of the highly-anticipated $\sqrt{s_{NN}} = 2.76$ TeV Pb-Pb collisions at the Large Hadron Collider (LHC) at the CERN, the heavy ion programs continue to explore the high energy frontiers of the QCD matter.

At the mid-high-energy collisions in the AGS, the SPS, and the early RHIC,  
the baryon stopping has been used to quantify the nuclear transparency in the collisions and the kinetic energy loss for the production of the hot medium. 
It is found that the average rapidity losses \cite{Busza:1983rj, Videbaek:1995mf} at the RHIC are apparently less than the simple linear scaling from the AGS \cite{Ahle:1999in} and the SPS \cite{Appelshauser:1998yb} above around $\sqrt{s_{NN}} = 62.4$ GeV \cite{Bearden:2003hx, Arsene:2009aa}, which suggests that the collisions would become increasingly transparent with the collision energy. The net baryon distribution itself would be an important observable in the higher energy collisions at the late RHIC and the LHC as shown in a number of theoretical analyses \cite{Bass:1999zq, Wolschin:1999jy,Soff:2002bn,Bass:2002vm,Itakura:2003jp,Pop:2004dq,Eskola:2005ue,Albacete:2006vv,Mishustin:2006wd,MehtarTani:2008qg,Ivanov:2010cu}, even though the net baryon is often treated as irrelevant and neglected because it becomes small near mid-rapidity as the collision energy increases. 
It is worth-mentioning that the net baryon number in the initial colliding nuclei is conserved throughout the time evolution and does not vanish from the system even at highest energies. The net baryon distribution might also carry valuable information on the valence quarks in the colliding nuclei and on the formation of the QGP itself in the early thermalization stage, which is not well understood yet.

It is quantitatively shown in the RHIC \cite{RHIC:summary} and the LHC \cite{Aamodt:2010pa} experiments that the relativistic hydrodynamic analysis is a powerful method to describe the dynamical behavior of the hot medium when its temperature is around and above the crossover temperature \cite{Kolb:2000fha,Teaney:2000cw,Hirano:2001eu}. The typical timescale for the hydrodynamic applicability is $\tau \sim 1$-$10$ fm/$c$. Recent progresses in hydrodynamic studies \cite{Schenke:2010rr,Bozek:2011ua,Vredevoogd:2012ui} incorporate non-equilibrium effects of shear and/or bulk viscosity into their calculations to quantitatively understand the properties of the QGP to explain particle spectra, azimuthal anisotropy, multiplicity, and rapidity distribution. 
On the other hand, the finite density effects are neglected in most of the modern hydrodynamic calculations with viscosity although they could be important in the context of precision physics. It is one of its great advantages that relativistic hydrodynamics can handle finite chemical potentials with ease even in off-equilibrium systems. The comparison of the experimental data to the hydrodynamical calculations of the net baryon distribution would, therefore, provide us with valuable insights for understanding the early dynamics as well as for determining the equation of state and the transport coefficients, such as baryon diffusion coefficient, at non-vanishing baryon density. The development of a consistent dissipative hydrodynamic model at finite density would also help the efforts to explore the QCD phase diagram in the search for the critical point. 

In this paper, I would like to estimate the effects of collective flow on the net baryon distribution and the average rapidity loss for the late RHIC and the LHC including the interplay of shear viscosity, bulk viscosity, and baryon dissipation. A relativistic dissipative hydrodynamic model for the longitudinal expansion of the quark matter is developed from a generalized version of the second order theory \cite{Israel:1979wp}. 
Here \textit{viscous} hydrodynamics refers to the hydrodynamics with shear and bulk viscosities, which are the tensor and the scalar off-equilibrium processes, and \textit{dissipative} hydrodynamics to the one with charge dissipations, the vector processes, in addition to the viscosites \cite{Monnai:2010qp}. 
The initial energy and the initial net baryon distributions are employed from the color glass theory \cite{McLerran:2001sr, Blaizot:2004px, Gelis:2010nm} and the equation of state at finite baryon density is constructed from lattice QCD results. 

The paper is organized as follows; Sec.~\ref{sec2} is devoted for the formulation of relativistic dissipative hydrodynamics with finite net baryon density. In Sec.~\ref{sec3}, the models for the equation of state and the transport coefficients for hydrodynamic calculations are introduced, along with initial conditions and freezeout. In Sec.~\ref{sec4}, I calculate rapidity distributions of the net baryon number to discuss the hydrodynamic effects with baryon dissipation as well as shear and bulk viscosities on the baryon stopping. Finally in Sec.~\ref{sec5}, discussion and conclusions are presented. The natural unit $c = \hbar = k_B = 1$ and the Minkowski metric $g^{\mu \nu} = \mathrm{diag}(+,-,-,-)$ is used throughout this paper.

\section{Relativistic Dissipative Hydrodynamics}
\label{sec2}
\vspace*{-2mm}

I develop a finite-density relativistic hydrodynamic model with off-equilibrium processes to describe the dynamical evolution of the quark-gluon plasma in high-energy heavy ion collisions. Baryon dissipation is taken into account for evolving flow for the first time, together with shear viscosity and bulk viscosity. A generalized version \cite{Monnai:2010qp} of the second order dissipative hydrodynamics \cite{Israel:1979wp} is introduced for the equations of motion for the dissipative currents to preserve causality and stability. Cross terms among different thermodynamic forces are present in the formalism. 
I focus on the longitudinal evolution of the target systems and integrate out the transverse dynamics \cite{Bozek:2007qt, Monnai:2011ju} in the study because the net baryon distribution is expected to depend mainly on rapidity as it is carried by the remnant of the valence quarks in the colliding nuclei at forward rapidity. It is also experimentally supported that the net baryon distributions do not depend on transverse geometry \cite{Adler:2003cb}. It should be emphasized that the longitudinal boost invariance \cite{Bjorken:1982qr} is not assumed.

The energy-momentum tensor $T^{\mu \nu}$ and the net baryon current $N_B^\mu$ are introduced as the conserving quantities in the system. In principle one can introduce other conserved currents such as the net strangeness current $N_s^\mu$, but those currents are expected to be much smaller than $N_B^\mu$ for the standard nucleus-nucleus collisions and are not considered here. 
The conserved quantities are related to the thermodynamic quantities through tensor decomposition as
\begin{eqnarray}
\label{eq:em_eqnoneq}
T ^{\mu \nu} &=& (e_0+\delta e) u^\mu u^\nu - (P_0+\Pi ) \Delta ^{\mu \nu} \nonumber \\
&+& 2W^{( \mu} u^{\nu )} + \pi^{\mu \nu} , \\
N_B^\mu &=& (n_{B0}+\delta n_B) u^\mu + V^\mu ,
\end{eqnarray}
where $u^\mu$ is the flow and $\Delta^{\mu \nu} = g^{\mu \nu}-u^\mu u^\nu$ is the projection operator perpendicular to the flow, \textit{i.e.}, $\Delta_{\mu \nu} u^\mu = 0$. In the local rest frame, $e_0$, $P_0$, and $n_{B0}$ are interpreted as energy density, hydrostatic pressure, and baryon number density. $\Pi$ is bulk pressure, $W^\mu$ energy dissipation current, $\pi^{\mu \nu}$ shear stress tensor, and $V^\mu$ baryon dissipation current. The distortions of energy and baryon number densities $\delta e$ and $\delta n_{B0}$ equal to zero because of the requirement of thermodynamic stability \cite{Monnai:2009ad}. 

The dissipative corrections in the linear response theory can be interpreted rather intuitively. The bulk pressure is a dynamical correction to the hydrostatic pressure which arises when the system is expanded/compressed without deformation, or when the temperature $T$ or the chemical potential $\mu_B$ decreases/increases. It is note-worthy that cancellation among the linear terms could be the reason for general smallness of the bulk pressure since the phenomena do not occur independently in hydrodynamic systems. See Appendix~\ref{appA} for the details. The shear stress tensor corresponds to the response to the deformation without volume change. The energy and the baryon dissipation currents are the local fluxes of energy and baryon densities which dissipate away from the flow, respectively.  They are induced by the spatial gradients in the temperature and the chemical potential. If one chooses the frame in the direction of the overall local energy flux, then $W^\mu = 0$ is concluded without the loss of generality. This frame is called the energy frame or Landau frame. From now on I consider this energy frame for the formulation of the hydrodynamic scheme. 

In (1+1)-dimensional dissipative hydrodynamics, the flow is expressed with the flow rapidity $Y_f$ as $u^\mu = (\cosh Y_f ,0,0, \sinh Y_f)$, where $z$-axis is the direction of the expansion. It follows from the orthogonality relation $\pi^{\mu \nu} u_\mu =0$ and the traceless condition $\pi^{\mu}_{\ \mu} = 0$ that the shear stress tensor can be expressed with a single independent variable $\pi = \pi^{00}-\pi^{33}$, which can be called \textit{shear pressure}, as
\begin{eqnarray}
\pi^{\mu \nu} = \left(
 \begin{array}{cccc}
  -\sinh^2{Y_f} & 0 & 0 & -\cosh{Y_f}\sinh{Y_f}\\
  0 & \frac{1}{2} & 0 & 0 \\
  0 & 0 & \frac{1}{2} & 0 \\
  -\cosh{Y_f}\sinh{Y_f} & 0 & 0 & -\cosh^2{Y_f} \\ 
 \end{array}
\right) \pi . \nonumber \\
\label{shear_pressure}
\end{eqnarray}
Likewise, because of the orthogonality relation $V^\mu u_\mu = 0$, the baryon dissipation current $V^\mu$ can be expressed as 
\begin{eqnarray}
V^\mu = (-\sinh{Y_f},0,0,-\cosh{Y_f}) V,
\label{baryon_dissipation}
\end{eqnarray}
where $V$ can be called \textit{baryon dissipation}.

The energy momentum and the net baryon number conservations then read
\begin{eqnarray}
D e_0 &=& -(e_0 + P_0 + \Pi - \pi ) \nabla Y_f  , \label{e} \\
(e_0 + P_0 + \Pi - \pi )D Y_f &=& -\nabla (P_0 + \Pi - \pi ) , \label{Yf} 
\end{eqnarray}
and
\begin{eqnarray}
D n_{B0} &=& - n_{B0} \nabla Y_f + \nabla V + V D Y_f, \label{nb} 
\end{eqnarray}
where the time- and the space-like derivatives in this geometry are
\begin{eqnarray}
D &=& \cosh (Y_f -\eta _s) \partial _\tau + \frac{1}{\tau} \sinh (Y_f -\eta _s) \partial _{\eta _s} , \label{D} \\
\nabla &=& \sinh (Y_f -\eta _s) \partial _\tau + \frac{1}{\tau} \cosh (Y_f -\eta _s) \partial _{\eta _s} .\label{K} 
\end{eqnarray}
Here $(\tau, \eta_s)$ is the relativistic coordinate defined as $t = \tau \cosh \eta_s$ and $z = \tau \sinh \eta_s$. $\tau$ is the proper time and $\eta_s$ the space-time rapidity.

One further needs three constitutive equations to determine the space-time evolution of the systems besides the equation of state $P_0 = P_0(e_0, n_{B0})$. Here I introduce the full second order dissipative hydrodynamic equations from Ref.~\cite{Monnai:2010qp} which extends the Israel-Stewart theory \cite{Israel:1979wp} for the systems with particle number changing processes. 
The constitutive equations for the bulk pressure $\Pi$, the baryon dissipation $V$, and the shear pressure $\pi$ in the (1+1)-dimensional geometry can be expressed as
\begin{eqnarray}
D \Pi &=& \frac{1}{\tau_\Pi} \bigg( -\Pi -\zeta_{\Pi \Pi} \frac{1}{T} \nabla Y_f - \zeta_{\Pi \delta e} D \frac{1}{T} +  \zeta_{\Pi \delta_{n_B}} D\frac{\mu_B}{T} \nonumber \\
&+& \chi_{\Pi \Pi} \Pi \nabla Y_f -\chi_{\Pi V}^{A} V \nabla \frac{\mu_{B}}{T} - \chi_{\Pi V}^B V \nabla \frac{1}{T} \nonumber \\
&+& \chi_{\Pi V}^C V D Y_f - \chi_{\Pi V}^D \nabla V + \chi_{\Pi \pi} \pi \nabla Y_f 
\bigg)
,
\label{eq:Pi}
\end{eqnarray}

\begin{eqnarray}
D V &=& \frac{1}{\tau _{V}} \bigg [ -V + \kappa_{V_B} \nabla \frac{\mu _B}{T} - \kappa _{V_J W} \bigg(\nabla \frac{1}{T} - \frac{1}{T} D Y_f \bigg) \nonumber \\
&+& \chi_{V V} V \nabla Y_f + \chi_{V \pi}^{A} \pi \nabla \frac{\mu _B}{T} + \chi_{V \pi}^B \pi \nabla \frac{1}{T} \nonumber \\
&-& \chi_{V_J \pi}^C \pi D Y_f \nonumber + \chi_{V \pi}^D \nabla \pi + \chi_{V \Pi}^{A} \Pi \nabla \frac{\mu _B}{T} \nonumber \\
&+& \chi_{V \Pi}^B \Pi \nabla \frac{1}{T} - \chi_{V \Pi}^C \Pi D Y_f + \chi_{V \Pi}^D \nabla \Pi \bigg] 
,
\label{eq:V}
\end{eqnarray}

\begin{eqnarray}
D \pi &=& \frac{1}{\tau_\pi} \bigg( -\pi + \frac{4}{3} \eta \nabla Y_f \nonumber \\
&+& \chi_{\pi \pi} \pi  \nabla Y_f  + \chi_{\pi \Pi} \Pi \nabla Y_f - \chi_{\pi V}^{A} V \nabla \frac{\mu_{B}}{T} \nonumber \\
&-& \chi_{\pi V}^B V \nabla \frac{1}{T} + \chi_{\pi V}^CV D Y_f - \chi_{\pi V}^D \nabla V \bigg) 
,
\label{eq:pi}
\end{eqnarray}
where $\eta$ is the shear viscosity, $\zeta_{\Pi \Pi}$, $\zeta_{\Pi \delta e}$, and $\zeta_{\Pi \delta n_B}$ the bulk viscosities, $\kappa_V$ the baryon charge conductivity and $\kappa_{VW}$ the baryon-heat cross conductivity. It should be noted that the linear cross terms, which implicitly satisfy Onsager reciprocal relations \cite{Onsager}, are explicitly present in the equations. The diagonal baryon charge conductivity is related with the baryon diffusion coefficient as $D_B = (\partial \mu_B / \partial n_{B0})|_T \kappa_V/T$. The cross conductivity causes the Soret effect, which is the chemical diffusion induced by the thermal gradient and flow acceleration. It is related to the thermo-diffusion coefficient as $D_T = (\partial \mu_B /\partial T)_{P_0} \kappa_V /n_{B0} + (\kappa_{VW} - \mu_B \kappa_V ) /T$. The ratio $k_T = D_T/D_B$ is called Soret coefficient. The cross coefficient can either be positive or negative while the semi-positive definite condition of the transport coefficient matrix $\kappa_{VW}^2 \leq \kappa_V \kappa_W$ is satisfied. Here $\kappa_W$ is the thermal conductivity but does not explicitly appear in the formalism because of the frame choice. 
$\tau_\Pi$, $\tau_\pi$ and $\tau_V$ are the relaxation times, $\chi_{\Pi \Pi}$, $\chi_{\pi \pi}$, and $\chi_{V V}$ the second order self-coupling coefficients, and $\chi_{\Pi \pi}$, $\chi_{\pi V}$, and $\chi_{V \Pi}$ the second order cross coefficients. The terms are combined using the conservation laws and the Gibbs-Duhem relation with the truncation to the second order. The couplings among different dissipative quantities at the second order can in principle have non-trivial effects when there are quantitative hierarchies among the dissipative quantities.

The equations involve many time-like derivatives, potentially increasing the numerical difficulties. Here they are solved with an advanced version of the multiple iteration algorithm I have developed for Ref.~\cite{Monnai:2011ju} coupled with the piece-wise parabolic method \cite{Colella:1982ee}.

\section{The Model}
\label{sec3}
\vspace*{-2mm}

The relativistic hydrodynamic model describes macroscopic motion of the fluid with the conservation laws and the constitutive equations. It means that the QCD equation of state and the QCD transport coefficients must be given as parameter input to perform hydrodynamic calculations for the quark matter. 
Also the hydrodynamic description works in the intermediate stage of the collisions around $\tau \sim 1$-$10$ fm/$c$. Thus one further needs to introduce initial conditions and freezeout to link hydrodynamic analyses with experimental observables. Here the color glass theory is employed to estimate the initial conditions for the energy density and the net baryon density profiles. The hydrodynamic flow is converted to particles spectra via freezeout.

\subsection{Equation of state and transport coefficients}

The equation of state (EoS) and the transport coefficients are the static and the dynamical responses of a thermodynamic system. They depend on microscopic properties of the medium, and are necessary input for hydrodynamic models.
Obtaining these quantities from the first principle calculations is, however, generally a very non-trivial issue, especially for finite baryon chemical potential systems due to the fermion-sign problem of the lattice QCD calculations.
Here the finite density EoS is constructed with the Taylor expansion method up to the second order,
\begin{eqnarray}
\frac{P(T,\mu_B)}{T^4} &=& \frac{P(T)}{T^4} + \frac{\chi_B^{(2)}(T)}{2} \bigg( \frac{\mu_B}{T} \bigg) ^2 + \mathcal{O} \bigg( \frac{\mu_B}{T} \bigg) ^4 ,
\label{eq:latticeeos}
\end{eqnarray}
where $\chi_B^{(2)}$ is the quadratic fluctuation of the baryon number. The latest (2+1)-flavor lattice QCD results of the continuum extrapolations for the EoS at vanishing chemical potentials \cite{Borsanyi:2010cj} and for the quadratic baryon fluctuation \cite{Borsanyi:2011sw} are employed.
It should be noted that the fugacity exponent $\mu_B/T$ has to be small for the expansion to be formally valid.
This is well motivated for the high-energy heavy ion collisions where the net baryon density is relatively small.

The transport coefficients of the hot matter are more difficult to obtain from the first principle calculations. To the best knowledge of the author, no conclusive results for these coefficients are available so far. This makes constraining the coefficients from experimental data in a dissipative hydrodynamic modeling one of the goals of heavy ion physics. Here I choose model coefficients for demonstrative purposes.
The shear viscous coefficient is introduced from the conjectured minimum boundary $\eta/s = 1/4\pi$ \cite{Kovtun:2004de} from Anti-de Sitter/conformal field theory (AdS/CFT) correspondences where $s$ denotes the entropy density.  
The three bulk viscous coefficients $\zeta_{\Pi \Pi}$, $\zeta_{\Pi \delta e}$, and $\zeta_{\Pi \delta n_B}$ are naturally expected in the linear response theory. $\zeta_{\Pi \Pi}$ is the diagonal component in the transport coefficient matrix and the others are off-diagonal, or cross, components. The ratios $\zeta/\eta$ are obtained from the non-equilibrium statistical operator method with a $\phi^4$-theory estimation \cite{Hosoya:1983id}. Following the discussion in Appendix~\ref{appA}, I combine them to \textit{effective} bulk viscous coefficient $\zeta = 5 ( \frac{1}{3} - c_s^2 ) \eta$ at the first order. The coefficient phenomenologically exhibit the peak structure around the crossover temperature $T_c$. 

\begin{figure}[tb]
\includegraphics[width=3.4in]{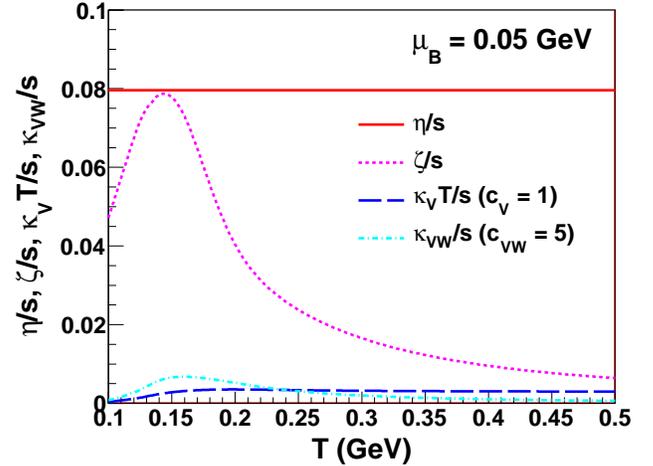}
\caption{(Color online) The dimensionless ratios of the shear viscosity (solid line), the effective bulk viscosity (dotted line), the baryon charge conductivity multiplied by temperature (dashed line), and the baryon-heat cross conductivity (dash-dotted line) to the entropy density at $\mu_B = 0.05$~GeV. The temperature region relevant to the numerical simulation $0.1$ GeV $\leq T \leq 0.5$ GeV is shown. }
\label{fig:1}
\end{figure}

The baryon charge conductivity and the baryon-heat cross coefficient characterize the finite density non-equilibrium processes. Using the charge conductivity $D_B = 1/2\pi T$ in an AdS/CFT framework \cite{Natsuume:2007ty}, the former is estimated as 
\begin{equation}
\kappa_V = \frac{c_V}{2\pi} \bigg( \frac{\partial \mu_B}{\partial n_B} \bigg)_T^{-1} = c_V \frac{\chi_B^{(2)} T^2}{2 \pi},
\end{equation} 
where the Taylor expansion of the EoS (\ref{eq:latticeeos}) is utilized. $c_V$ is a dimensionless constant introduced for parametrization. Here $c_V = 1$ is considered unless specified otherwise.
The cross coefficient is parametrized from the dimensional analyses, the matter-antimatter symmetry $V^\mu (\mu_B) = -V^\mu (-\mu_B)$, and the implication from the semi-positive definiteness as $\kappa_{VW} = c_{VW} [n_{B0} T/(e_0+P_0)] \sqrt{\kappa_V \kappa_{W}}$.
The semi-positive definite condition of the transport coefficient matrix in this case is explicitly expressed as $c_{VW}^2 [n_{B0} T/(e_0+P_0)]^2 \leq 1$. 
Here the thermal conductivity is chosen as $\kappa_W = 5 \eta T$ \cite{Hosoya:1983id}.
Note that the charge conductivity in this model remains finite in the limit of vanishing chemical potential, but does not induce the charge dissipation current out of global chemical equilibrium because the cross coefficients vanish in the limit and it forms an isolated partial matrix in the full transport coefficient matrix. The lack of cross coefficients at the vanishing limit of the corresponding chemical potentials is important because it ensures that hidden conserving quantities -- strangeness or yet unknown charge -- do not affect the physics.
The dimensionless ratios of the linear transport coefficients $\eta/s$, $\zeta/s$, $\kappa_V T/s$, and $\kappa_{VW} /s$ with $c_V = 1$ and $c_{WV} = 5$ for a constant chemical potential $\mu_B = 0.05$~GeV are plotted in Fig.~\ref{fig:1}. One can see that the baryon-related transport coefficients are smaller than the others in the current parameter settings. 

The relaxation times are also estimated from the string theoretical framework as $\tau_\pi = (2 - \ln 2) / 2\pi T$, $\tau_\Pi = 18 - (9\ln 3 - \sqrt{3} \pi) / 24\pi T$, and $\tau_V = \ln 2 / 2\pi T$ \cite{Natsuume:2007ty}. The other second order transport coefficients are parametrized as $\chi_{AB} = \tilde{c}_{AB} [n_{B0} T/(e_0+P_0)]^n T^m \tau_A$ where $A$ and $B$ denote the types of dissipative currents and $\tilde{c}_{AB}$ is a dimensionless constant. 
$n = 1$ is employed for the baryon-non-baryon cross coefficients and $n = 0$ for the others, following the matter-antimatter symmetry arguments mentioned earlier. $m$ is the dimension parameter chosen so that the dimensions of the terms in each constitutive equation are matched. For the most part in this study, however, $\tilde{c}_{AB} = 0$ is employed to observe the qualitative nature of the dissipative processes. Note that, in the present analyses, the focus is on the net baryon density and the interplay of different dissipative currents, and it is beyond the scope of the paper to precisely obtain the transport coefficients.

\subsection{Initial conditions}

The initial condition for the energy density is constructed in the color glass theory. The theory describes the colliding two nuclei as saturated gluons called color glass condensate (CGC). Since the gluons are dominant against the valence quarks, the initial energy density of the hot matter is estimated from the gluon distribution by assuming the profile of the energy distribution is not significantly modified during the early thermalization stage. The Nara Monte-Carlo adaptation \cite{Hirano:2004en,Drescher:2006pi,Drescher:2006ca,Drescher:2007ax} of Kharzeev-Levin-Nardi model (MC-KLN) \cite{Kharzeev:2002ei} is employed for the estimation. 
The rapidity distribution of the transverse gluon energy density can be expressed in $k_T$-factorization as
\begin{eqnarray}
\frac{dE_T}{d^2 r_T dy} &=& \frac{4\pi N_c \alpha_s}{N_c^2-1} \int \frac{d^2 p_T}{p_T} \int d^2 k_T \varphi_1 (x_1, k_T^2) \nonumber \\
&\times& \varphi_2 (x_2,(p_T-k_T)^2) ,
\end{eqnarray}
where $N_c$ = 3 is the number of colors, $\alpha_S$ the QCD coupling, $x_{1,2} = p_T \exp (\pm y) /\sqrt{s}$, and $\varphi_{1,2}$ the unintegrated gluon distributions. The lower index $T$ denotes the transverse component of given position or momentum. The gluon distribution is saturated at the scale $Q_{s,A}^2 (x, r_\perp)= Q_{s,0}^2 [T_A(r_\perp)/T_{A,0}] (x_0/x)^\lambda $ where $\lambda = 0.28$ is experimentally motivated. $T_A$ is the thickness function to account for the nucleus geometry. Here the parameters are chosen as $x_0 = 0.01$, $T_{A,0} = 1.53$ fm$^{-2}$, and $Q^2_{s,0} = 2$ GeV \cite{Drescher:2006pi}. Since the hydrodynamic medium is expected to be constituted by low $p_T$ partons, the $p_T$ window $0.1$-$3.0$ GeV is set. The space-time rapidity $\eta_s$ is matched with the momentum rapidity $y$ to obtain the initial condition in the configuration space. 

The initial condition for the net baryon density is also constructed from the color glass picture by assuming it is proportional to the valence quark parton distribution function \cite{MehtarTani:2008qg}. The net baryon distribution for a nucleus reads
\begin{eqnarray}
\frac{dN_{B-\bar{B}}}{dy} = \frac{C}{(2\pi)^2} \int \frac{d^2 p_T}{p_T^2} x_1 q_v (x_1) \varphi (x_2, p_T) ,
\end{eqnarray}
where $q_v$ is the valence quark distribution. Here the normalization $C$ is determined so that the integrated $N_{B-\bar{B}}$ matches the number of participants. The distributions are determined with the same settings as the no-fragmentation case in Ref.~\cite{MehtarTani:2008qg} but with $\lambda = 0.28$ and the next-to-next-to-leading order fit results of the valence quark distribution \cite{Martin:2002dr}. 
Note that the net baryon distribution is sensitive to the parameters, and here they are chosen to yield much steeper distribution at the initial stage so the distribution after the hydrodynamic evolution roughly reproduces experimental data.
The tail contribution beyond the beam rapidity is exponentially cut off since hydrodynamic description do not apply to the region beyond the freezeout and the Taylor expansion-based EoS is not expected to work for the dense and cold matter. This leads to the reduction of the total baryon number of participant nucleons by 8.7\% at the RHIC and 0.1\% at the LHC for the most central 0-5\% events. 

The initial conditions for the dissipative currents are not well known and this is also a quite interesting issue by itself. Here they are chosen as non-existent at the initial time, \textit{i.e.}, $\Pi (\tau_0, \eta_s) = V (\tau_0, \eta_s) = \pi (\tau_0, \eta_s) = 0$ for the clear view of the non-equilibrium effects on the fluids and also for avoiding possible overestimation of the viscous and the dissipative effects from ambiguity. Here $\tau_0$ is the initial time set to $\tau_0 = 1$ fm/$c$.

\subsection{Freezeout}

As the system cools down with the time evolution and becomes dilute enough, the hydrodynamic simulation has to be stopped and the flow field needs to be converted into particles. One conventionally employs the Cooper-Frye formula \cite{Cooper:1974mv} at a freezeout hypersurface $\Sigma$. The formula reads 
\begin{equation}
\label{eq:spectra}
\frac{d^2N_i}{d^2p_Tdy} = \frac{g_i}{(2\pi )^3} \int _\Sigma p_i^\mu d \sigma _\mu f_i,
\end{equation}
for the particle species $i$, where $g_i$ is the degeneracy, $d \sigma _\mu$ the freezeout hypersurface element, and $f_i$ the phase-space distribution. $\Sigma$ is taken as an isothermal surface because the chemical potential dependence of the boundary is sufficiently small for the high-energy collisions. This gives rise to the concept of the freezeout temperature $T_f$. $f_i$ can be separated into the equilibrium distribution $f_i^0$ and the distortion of distribution $\delta f_i$, where
\begin{equation}
\label{eq:f0}
f_i^0 = \bigg[ \exp{\bigg( \frac{p^i_\mu u^\mu - b_i \mu_B }{T}\bigg) } - \epsilon_i \bigg]^{-1}.
\end{equation}
Here $b_i$ is the baryon number, \textit{i.e.}, $b_i = +1$ for baryons, $-1$ for anti-baryons, and $0$ for mesons. $\epsilon_i$ denotes the quantum statistics as $\epsilon_i = +1$ for fermions and $-1$ for bosons.  The effects of the off-equilibrium distribution $\delta f_i$ is not treated here because 
its correction on the $p_T$-integrated net baryon rapidity distribution would be small while the off-equilibrium expansion is applicable, since the stability condition for the net baryon density requires $\delta n_B = 0$ and $|\delta N_B^\mu|/|N_{B0}^\mu| = V/n_{B0}$ would be generally very small. 

It should be emphasized here that the focus of this paper is to estimate the finite-density hydrodynamic effects with non-equilibrium processes on the hot medium because it would have the dominant effects on the net baryon distribution. The hadronic cascade at the later stage is not considered because the modification on the net baryon distribution during the hadronic stage would be small as the baryon number does not change in the hadronic decay and diffusion process in a hadronic gas is expected to be slow \cite{Bass:2002vm, Shuryak:2000pd}. 

The information on the chemical freezeout is implicitly contained in the lattice QCD EoS in the hadronic phase since the hydrodynamic flow does not specify its contents. If the EoS for each hadronic component was known, one could in principle incorporate the chemical freezeout explicitly by introducing conservation laws for the hadrons instead of the one for the baryon charge \cite{Teaney:2001av}.

\section{Results}
\label{sec4}
\vspace*{-2mm}

The most central 0-5\% events are considered for the initial conditions. The mean numbers of participants are 357 for the RHIC and 385 for the LHC.
The initial temperatures and the chemical potentials at mid-rapidity are $T=419$ MeV and $\mu_B = 20.6$ MeV for the RHIC and $T=490$ MeV and $\mu_B = 6.5$ MeV for the LHC. 
I employ two freezeout temperatures $T_f = 0.16$ GeV and $T_f = 0.14$ GeV. The former early freezeout scenario is motivated by the possible break-down of hydrodynamic applicability due to the bulk viscous effects \cite{Torrieri:2007fb}. 
The net baryon distributions are constructed by taking into account the contributions of the hadron resonance \cite{Nakamura:2010zzi} up to $2.5$ GeV at freezeout.

\subsection{Net baryon distributions at the RHIC}

\begin{figure}[tb]
\includegraphics[width=3.4in]{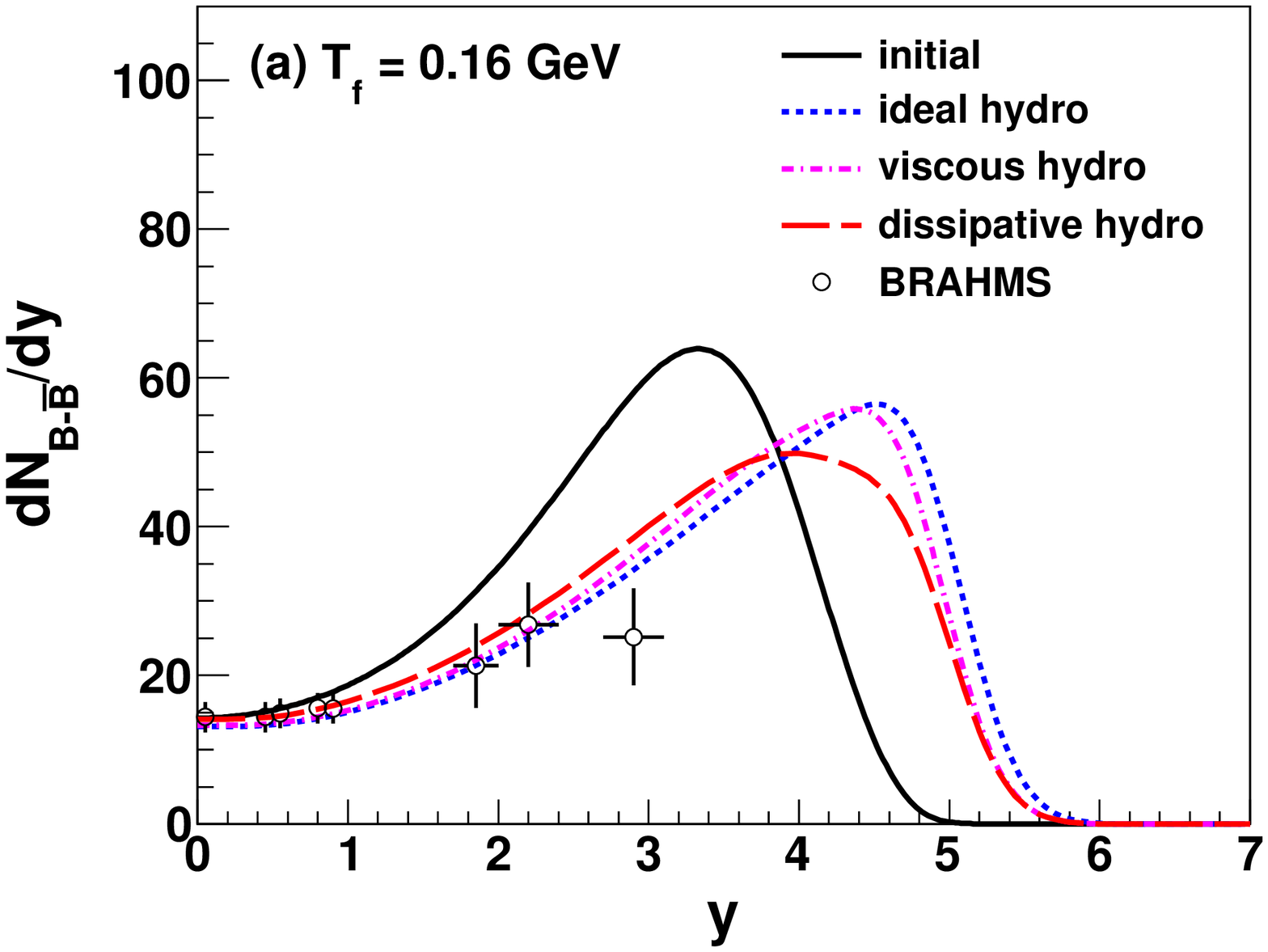} \\
\includegraphics[width=3.4in]{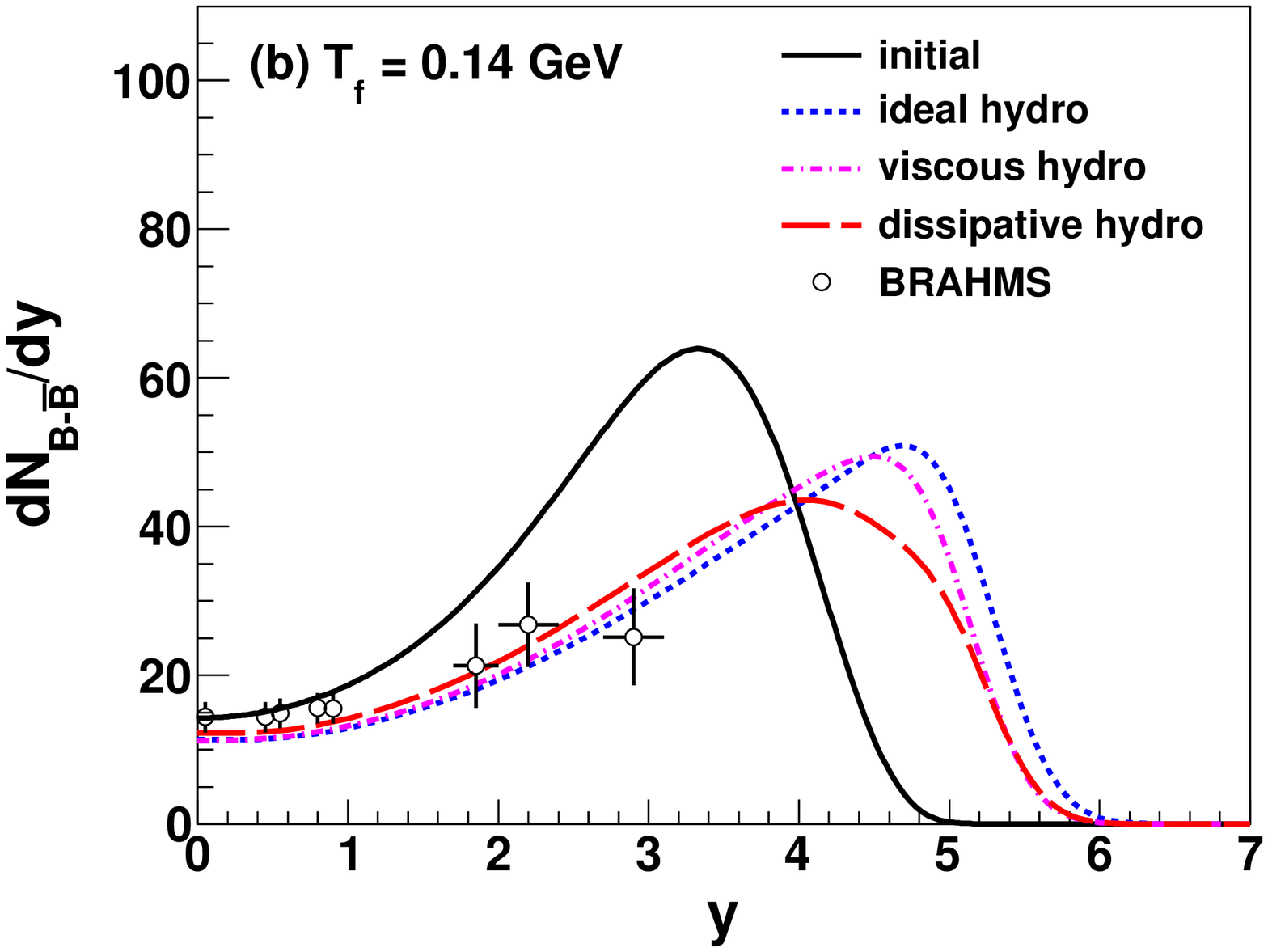}
\caption{(Color online) The initial net baryon distributions based on the color glass theory (solid curve) and the ones with the ideal (dotted curve), the viscous (dash-dotted curve), and the dissipative (dashed curve) hydrodynamic effects at (a) $T_f = 0.16$ GeV and (b) $T_f = 0.14$ GeV for the Au-Au collisions at $\sqrt{s_{NN}} = 200$ GeV. The experimental data points are the scaled results of the net proton distribution from the BRAHMS collaboration \cite{Bearden:2003hx}.}
\label{fig:2}
\end{figure}

The net baryon distributions of the Au-Au collisions at $\sqrt{s_{NN}} = 200$ GeV with and without non-equilibrium corrections are shown in Fig.~\ref{fig:2} for the freezeout temperatures $T_f = 0.16$ GeV and $T_f = 0.14$ GeV. The off-equilibrium parameters are set as $c_V = 1$, $c_{VW} = 0$, and $\tilde{c}_{AB} = 0$ for the moment. One can see that the hydrodynamic flow tends to carry the net baryon density to forward rapidity, broadening the flat region with relatively small baryon density around mid-rapidity. The viscous hydrodynamic results are less flattened because the shear and the bulk viscosities prevent the expansion by effectively reducing the longitudinal pressure. The dissipative hydrodynamic results, which include the baryon dissipation in addition to the viscosities, differ from the viscous hydrodynamic results. The gradient in the chemical potential induces the baryon dissipation current into mid-rapidity region, further steepening the mid-rapidity valley. 
The data points are scaled from the net proton distribution of the BRAHM experiments \cite{Bearden:2003hx}, which show relatively good agreement with the hydrodynamic results. It should be noted that quasi-quantitative discussion is allowed here because the initial distribution is sensitive to the parameters in the color glass theory and also the transverse dynamics is integrated out.

The magnitude of collisional transparency is quantified by the baryon stopping.
The rapidity loss is defined as $\langle \delta y \rangle = y_p - \langle y \rangle$ where $y_p$ is the rapidity of incoming projectile and 
\begin{equation}
\label{eq:meany}
\langle y \rangle = \int _0^{y_p} y \frac{dN_{B-\bar{B}}(y)}{dy} dy  \bigg/ \int _0^{y_p} \frac{dN_{B-\bar{B}}(y)}{dy} dy.
\end{equation}
The ideal, the viscous, and the dissipative hydrodynamic evolutions lead to the reduction in the average rapidity loss to $\langle \delta y \rangle$ = 2.09, 2.16, and 2.26 for $T_f = 0.16$ GeV and $\langle \delta y \rangle$ = 1.99, 2.06, and 2.19 for $T_f = 0.14$ GeV, respectively, when that of the initial net baryon distribution is $\langle \delta y \rangle$ = 2.67. 
The fact that hydrodynamic evolutions visibly reduce the rapidity loss suggests that the baryon stopping in the RHIC would deviate \textit{less} significantly from the linear extrapolation of the AGS and the SPS results at the formation of the hot medium, but the net baryon is carried to the forward rapidity by the hydrodynamic medium interaction, effectively enhancing the observed transparency. It also indicates that the kinetic energy loss for the QGP production is larger, and part of the energy is transferred back to the net baryon component from the produced medium afterward.

The fact that the effects of the baryon dissipation current could be visible on the net baryon distribution is of importance because it suggests that one would have to take the diffusion process into account to quantitatively understand the experimental data. It would also play an important role in constraining the yet unknown initial condition of the net baryon distribution. The actual effect of the baryon diffusion could be larger because the current baryon charge conductivity is moderate as shown in Fig. 1. It is note-worthy that though the shear and bulk viscosity and the baryon dissipation seem to have similar effects on the net baryon distribution, the former enhances the baryon and the anti-baryon distributions individually while the latter only increases their difference, suggesting that its contribution to the averaged distributions is small. 

Comparing the two freezeout temperatures, the broadening effect is larger for the late freezeout case because of the longer hydrodynamic evolution. This indicates that the late freezeout tends to allow larger dissipative and viscous coefficients. It should be noted that the total baryon number is slightly smaller than the initial nuclei at $T_f = 0.14$ TeV because of the Cooper-Frye formulation of the freezeout. The equation of state and the baryon fluctuation of the kinetic theory and those of the lattice QCD have to be identical at freezeout to perfectly conserve the energy and the net baryon because they are to be reproduced in relativistic kinetic theory from the flow, the temperature, and the chemical potential. This would make it slightly more difficult to distinguish the freezeout temperature dependence from the slopes of the net baryon distribution, but of course leaves the freezeout temperature dependence of the average rapidity losses unaffected. 

\subsection{Effects of cross terms}

\begin{figure}[tb]
\includegraphics[width=3.4in]{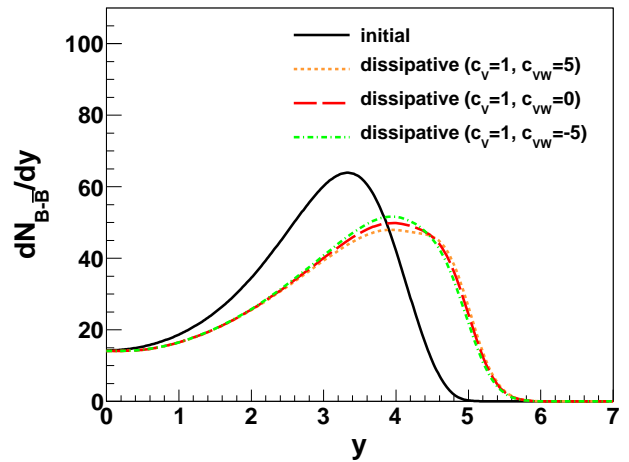} 
\caption{(Color online) The net baryon distribution at the initial stage based on the color glass theory (solid curve) and the ones after dissipative hydrodynamic evolutions with the cross coefficients $c_{VW} = 5$ (dotted curve), $c_{VW} = 0$ (dashed curve), and $c_{VW} = -5$ (dash-dotted curve) at $T_f = 0.16$ GeV for the Au-Au collisions at $\sqrt{s_{NN}} = 200$ GeV. }
\label{fig:3}
\end{figure}

I next explore the effects of the cross terms in the equations of motion for the dissipative currents (\ref{eq:Pi})-(\ref{eq:pi}). Fig.~\ref{fig:3} shows the Soret effect in the QGP medium, which is induced by the linear thermo-diffusion term in the baryon dissipation, at the RHIC for $T_f = 0.16$ GeV. $c_{VW} = 5, 0$, and $-5$ are employed for the cross coefficient. The semi-positive definite condition is checked to be satisfied throughout the time evolution. One can see that the positive and the negative cross coefficients qualitatively lead to reduction and enhancement of the baryon diffusion effect, respectively. The results are consistent with the linear order analyses in Eq.~(\ref{eq:vector_relation}) that the positive thermo-diffusion coefficient effectively reduces the charge conductivity. The magnitude of the thermo-diffusion effect, on the other hand, is relatively small in the present calculations because the cross coefficient becomes non-vanishing only in baryon rich region at forward rapidity in high-energy heavy ion collisions. This suggests that Soret effect might change the temperature and the chemical potential dependences of the transport properties, but it would be effective only at forward rapidity.

The second order cross terms could also affect the results because of the hierarchy in the magnitude of dissipative currents mentioned in Sec.~\ref{sec2}. Numerical estimations with finite $\tilde{c}_{AB}$ indicate that the bulk-shear cross term in the bulk pressure, and the baryon-shear and the baryon-bulk terms in the baryon dissipation can be relevant, assuming the magnitude of the transport coefficients is roughly of the same order. Note that in general there is much ambiguity in the magnitude of the second order transport coefficients. The result is consistent with the fact that the shear pressure is larger than the bulk pressure, which in turn is larger than the baryon dissipation in relativistic heavy ion collisions.  

\subsection{Net baryon distributions at the LHC}

\begin{figure}[tb]
\includegraphics[width=3.4in]{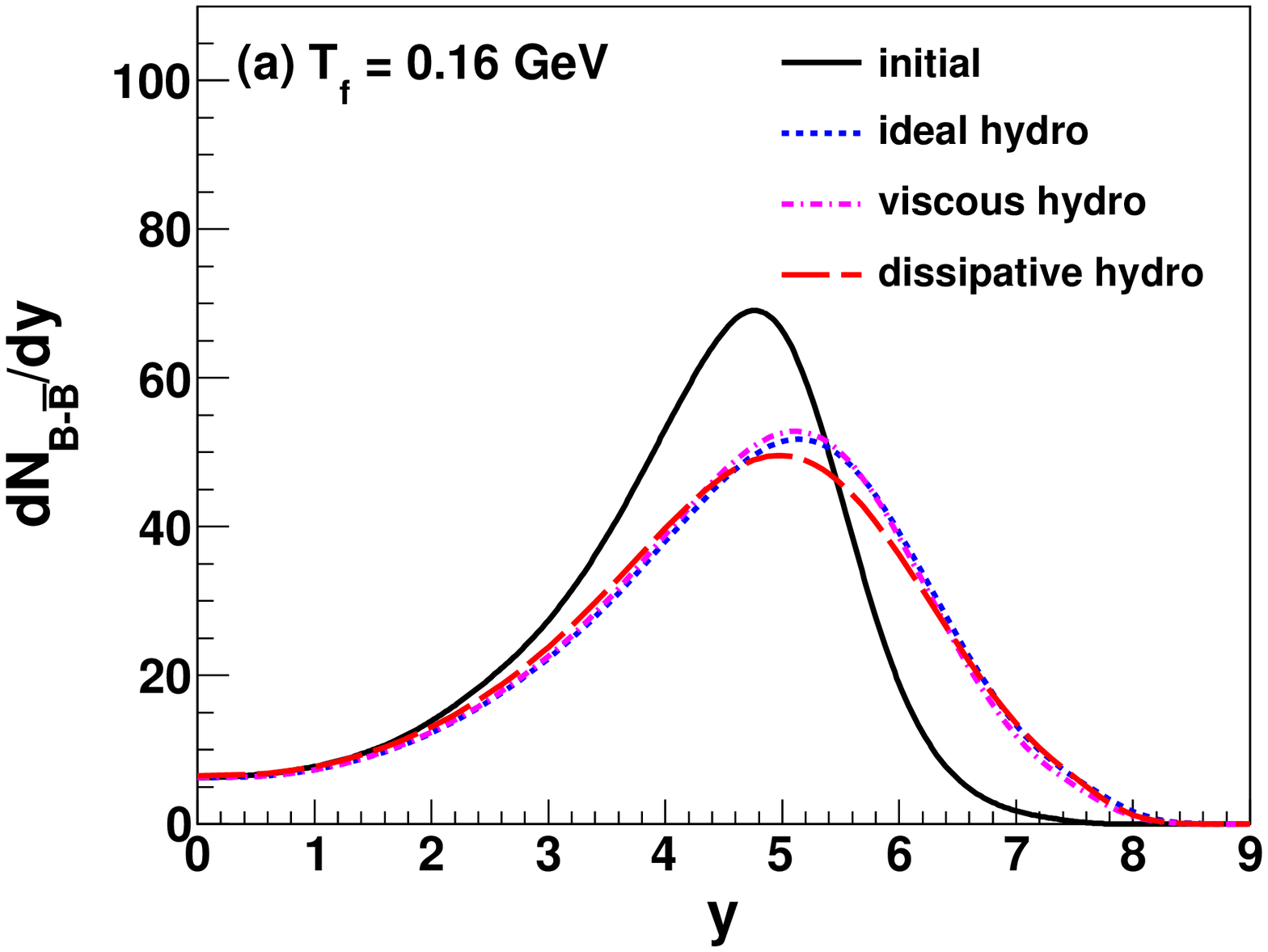}\\
\includegraphics[width=3.4in]{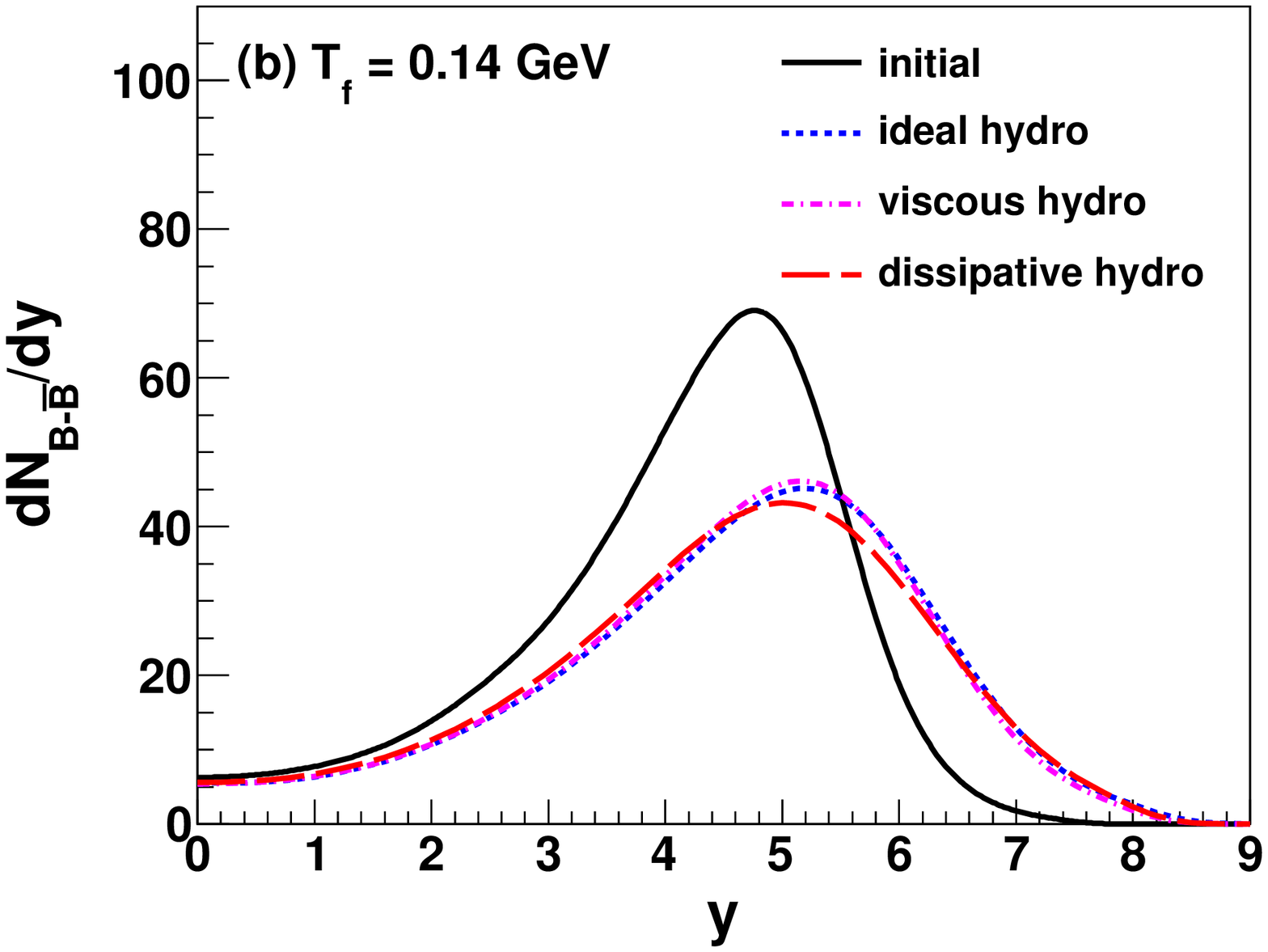}
\caption{(Color online) The initial net baryon distributions based on the color glass theory (solid curve) and the ones with the ideal (dotted curve), the viscous (dash-dotted curve), and the dissipative (dashed curve) hydrodynamic effects at (a) $T_f = 0.16$ GeV and (b) $T_f = 0.14$ GeV for the Pb-Pb collisions at $\sqrt{s_{NN}} = 2.76$ GeV.}
\label{fig:4}
\end{figure}

The prospects for the Pb-Pb collisions at $\sqrt{s_{NN}} = 2.76$ TeV in the LHC experiment are shown in Fig.~\ref{fig:4}. The net baryon distributions are still visibly modified by the hydrodynamic flow, effectively increasing the nuclear transparency in the collision. On the other hand, the effects of viscosities and dissipation are much smaller. The finial average rapidity losses after the ideal, the viscous, and the dissipative hydrodynamic evolution are $\langle \delta y \rangle$ = 3.48, 3.52, and 3.55 for $T_f = 0.16$ GeV and $\langle \delta y \rangle$ = 3.44, 3.48, and 3.51 for $T_f = 0.14$ GeV, respectively. The initial rapidity loss in the current parameter settings is $\langle \delta y \rangle$ = 3.88.  Comparing the early and the late freezeout cases, the latter tends to leave more room for viscosities and dissipation as is found in the RHIC settings.  

The smaller dissipative effect could be understood as a result of the smaller spatial gradients in the fugacity exponent $\mu_B/T$, the thermodynamic force to the baryon dissipation, at the LHC. The shear and the bulk viscous effects are reduced for a different reason, because they do not directly respond to the difference in the fugacity exponent. The hydrostatic pressure $P_0$ increases with the collision energy more than the shear and the bulk pressures do, and the effects of viscous corrections are decreased in the effective pressure $P = P_0 + \Pi - \pi$, reducing the difference between the ideal and the viscous hydrodynamic results. It should be noted that the effect of baryon dissipation could be larger since the charge conductivity employed in the estimations is very moderate as mentioned earlier. Also the off-equilibrium corrections on the net baryon distribution might be underestimated due to the lack of explicit chemical potential dependence of the transport coefficients including the shear and the bulk viscosities in the current modeling. 

Comparing the results with the ones at the RHIC, one can also find that the overall hydrodynamic effect is smaller at the LHC. One of the reasons would be the fact that the mean rapidity loss in the initial net baryon distribution for the current parameter settings is large, \textit{i.e.}, the peak of the distribution is around $y \sim 4$-$5$ while the beam rapidity is $y \sim 8$. The pressure is larger at the LHC, but the pressure gradient, which drives the net baryon current, is relatively small in this rapidity region. Thus the actual hydrodynamic effect could be large enough to be measured in the LHC experiments for the initial distributions with smaller mean rapidity loss or larger pressure gradient.

\section{Discussion and Conclusions}
\label{sec5}
\vspace*{-2mm}

A relativistic dissipative hydrodynamic model of the high-energy heavy ion collisions at finite density which takes account of shear viscosity, bulk viscosity, and baryon dissipation with evolving flow is developed for the first time. The hydrodynamic framework is employed from the second order theory extended for the systems with particle number changing processes. The initial conditions for the energy and the net baryon distributions are constructed from the color glass theory, and the EoS is employed from the Taylor expansion approach of the finite density lattice QCD to improve quantitative accuracy. I find that the average rapidity loss for the baryon stopping is reduced during the hydrodynamic evolution, which would mean that the observed transparency of the collision at the RHIC is effectively enhanced in the medium interaction. This suggests that more energy is available for the production of the hot medium at the initial stage than was implied from the experimental data, and the strongly-coupled medium re-distribute part of the energy back to the net baryon components.

The net baryon distribution could also be sensitive to baryon dissipation as much as to viscosities. It should be noted that the current dissipative coefficients are rather moderate as shown in Fig.~\ref{fig:1} and the actual diffusion effects can be larger. The effect of baryon dissipation could be important for explaining the experimental data and constraining the initial conditions, which are not well known. One would need to introduce other observables such as the transverse momentum spectra of the net baryon to constrain the transport coefficients of the hot QCD matter at finite baryon density from the collider experiments because of the ambiguities in the choice of initial conditions. The effects of the cross coefficients are also numerically investigated, and it is found that the thermo-diffusion effect, or the Soret effect, might modify the magnitude of the baryon diffusion, but the effect is limited to the forward rapidity region because the baryon-heat cross conductivity vanishes for the baryon-free medium due to the matter-antimatter symmetry. The second-order cross terms among the dissipative currents of different magnitudes would also be important for the quantitative analyses. The late freezeout is found to allow larger viscosities and baryon dissipation since the longer hydrodynamic evolution widens the mid-rapidity valley.

A possible source of overestimation for the hydrodynamic effects would be the lack of transverse expansion. The temperature and the chemical potential tend to be larger for the longitudinal geometry as the energy and the net baryon densities cannot spread into the transverse directions. It should be noted that while the transverse dependence of the net baryon distributions is experimentally implied to be small \cite{Adler:2003cb}, the accelerated cooling would lead to the reduction in the effect of hydrodynamic evolution. This is partially taken into account by employing the early freezeout scenario. The off-equilibrium corrections with transverse expansion would be more non-trivial and worth investigating. Also the parametrization of the transport coefficients for finite density systems, especially the conductivities $\kappa_V$ and $\kappa_{VW}$, needs to be improved through theoretical and experimental developments for more quantitative discussion. 

\begin{acknowledgments}
The author is grateful for the valuable comments by T.~Hatsuda. 
The work of A.M. is supported by JSPS Research Fellowships for Young Scientists.
\end{acknowledgments}

\appendix
\section{TRANSPORT COEFFICIENTS AT THE LINEAR ORDER}
\label{appA}
\vspace*{-2mm}

The determination of the transport coefficients for a given system has been one of the long standing issues even at the linear order. The bulk pressure at the first order is expressed as
\begin{eqnarray}
\Pi &=& - \zeta_{\Pi \Pi} \frac{1}{T}  \nabla_\mu u^\mu - \zeta_{\Pi \delta_e} D \frac{1}{T} + \zeta_{\Pi \delta n_B} D \frac{\mu_B}{T} \nonumber \\
&=& - \bigg[ \frac{ \zeta_{\Pi \Pi}}{T} + \frac{\zeta_{\Pi \delta_e}}{T} \bigg( \frac{\partial P_0}{\partial e_0} \bigg)_{n_{B0}} + \frac{\zeta_{\Pi \delta n_B}}{T} \bigg( \frac{\partial P_0}{\partial n_{B0}} \bigg) _{e_0} \bigg] \nonumber \\
&\times& \nabla _\mu u^\mu + \mathcal{O}(\delta^2) \nonumber \\
&\equiv& - \zeta \nabla _\mu u^\mu + \mathcal{O}(\delta^2) .
\label{eq:scalar_relation}
\end{eqnarray}
A naive $\phi^4$-theory analysis \cite{Hosoya:1983id} yields $\zeta_{\Pi \delta e} = - 3 \zeta_{\Pi \Pi}$ along with $\zeta_{\Pi \Pi}/T = 5 \eta/3$. If the system is free of conserved charge currents, the energy-momentum conservation and the Gibbs-Duhem relation yield
\begin{equation}
\zeta = 5 \bigg( \frac{1}{3} - c_s^2 \bigg) \eta ,
\label{eq:zeta}
\end{equation}
which satisfies the conjectured lower boundary $\zeta \geq 2 \big( \frac{1}{3} - c_s^2 \big) \eta$ in the $\mathcal{N}=2^{*}$ gauge theory \cite{Buchel:2007mf}. Since the squared sound velocity is around $1/3$ except for the crossover regions, the cancellation of the linear terms lead to small bulk viscosity. It implies that the existence of the Onsager cross term is a reason for the general smallness of bulk viscous coefficient in hydrodynamic systems. In other words, the diagonal bulk viscosity $\zeta_{\Pi \Pi}$ prevents the expansion of a system but the energy cross coefficient $\zeta_{\Pi \delta e}$ encourages the expansion in the effort of decreasing the temperature, leading to the overall cancellation of the effects.

When the system has finite chemical potential, the expression is subject to non-trivial contribution from the density cross coefficient. For a special case where $\zeta_{\Pi \delta {n_B}} = [n_{B0}/(e_0+P_0)] \times \zeta_{\Pi \delta e}$, however, the expression is dramatically simplified and one again obtains Eq.~(\ref{eq:zeta}), because the sound velocity is expressed as
\begin{eqnarray}
c_s^2 &=& \bigg( \frac{\partial P_0}{\partial e_0} \bigg)_{s/n_{B0}} \nonumber \\
&=& \bigg( \frac{\partial P_0}{\partial e_0} \bigg)_{n_{B0}} + \frac{n_{B0}}{e_0 + P_0} \bigg( \frac{\partial P_0}{\partial n_{B0}} \bigg) _{e_0} .
\label{eq:sv}
\end{eqnarray}
The sign of $\zeta_{\Pi \delta{n_B}}$ flips for antimatter systems, correctly capturing the fact that the overall bulk viscosity $\zeta$ is symmetric under the charge conjugation. I employ this form of bulk viscous coefficient since the main focus is on the net baryon density and its dissipation.

The baryon dissipation current can be expressed in a similar fashion using the conservation laws and the thermodynamic relation as 
\begin{eqnarray}
V^\mu &=& \kappa_{V} \nabla^\mu \frac{\mu_B}{T} - \kappa_{VW} \bigg( \nabla^\mu \frac{1}{T} + \frac{1}{T} D u^\mu \bigg) \nonumber \\
&=& \bigg( \kappa_V - \frac{n_{B0}}{e_0+P_0} \kappa_{VW} \bigg) \nabla^\mu \frac{\mu_B}{T} + \mathcal{O}(\delta^2) \nonumber \\
&\equiv& \kappa \nabla^\mu \frac{\mu_B}{T}  + \mathcal{O}(\delta^2) ,
\label{eq:vector_relation}
\end{eqnarray}
where $\kappa$ is the effective baryon charge conductivity. An extreme case is $\kappa_V = [n_{B0}/(e_0+P_0)] \times \kappa_{VW}$, where the current vanishes altogether at this order because of the Soret effect. It should be noted that the vector cross terms arising from multiple conserved currents cannot be integrated in this method. In the paper, the two coefficients are treated separately to better illustrate the role of each linear term. 

\bibliography{basename of .bib file}

\end{document}